\documentclass[12pt]{article}

\usepackage{graphicx}

\begin{document}
\begin{center}
{\Large\bf \boldmath  The Knight of the Quantum: On the Contribution of 
D.I. Blokhintsev to Quantum Physics}\footnote{Plenary Talk, given at XIII-th Int. Conf. on Selected Problems
of Modern Theoretical Physics, Dubna, 22--27 June 2008} \\ 
 
\vspace*{6mm}
{A. L. Kuzemsky   }\\      
{\small \it  Joint Institute for Nuclear Research, Dubna, Russia\\
kuzemsky@theor.jinr.ru;   http://theor.jinr.ru/\symbol{126}kuzemsky}         
\end{center}

\vspace*{6mm}

\begin{abstract}
A concise survey of the contribution of D.I. Blokhintsev to the quantum physics, including
solid state physics, physics of metals, surface physics, statistical physics and optics is given.
These achievements have been considered in the context of modern development of these fields of physics.
\end{abstract}

\vspace*{6mm}

The name of Corresponding Member of the Academy
of Sciences of the USSR D. I. Blokhintsev (January
11, 1908 -- January 27, 1979) is widely known in Russia
and abroad. His books are being republished; information
on his biography and his scientific heritage can
be found in multiple papers and collections of papers. 
However, for many scientists, his name is related
mainly to his works in the field of atomic and nuclear
physics, applied acoustics, participation in the creation
of the first nuclear power station in Obninsk, reactor
construction, and multiple studies in high energy and
elementary particle physics. It is not so well known that
at first he wrote some quite interesting and important
works in the field of quantum solid state physics and
statistical physics. 
In the beginning of his distinguished academic career~\cite{dib1,vrich}, D.I. Blokhintsev  has worked in the
field of  quantum solid state physics and statistical physics, as well as in the field of quantum physics~\cite{dib1}. 
The aim of my talk is to recall these quite interesting and important works and correlate them with corresponding modern directions in condensed matter 
physics and quantum physics~\cite{kuzrev}. D.I. Blokhintsev entered the Physics Faculty of Moscow State University in 1926.
At that time L.I. Mandelstam was the head of the Department of theoretical physics and optics and I.E. Tamm was professor of
theoretical physics of that Department. Blokhintsev considered L.I. Mandelstam, S.I. Vavilov and I.E. Tamm his teachers.
I.E. Tamm become his Ph.D. promotor in postgraduate studies. 
Thus, Blokhintsev's student years brought him
great and fruitful experience in communicating, at lectures
and in laboratories, with brilliant and interesting
representatives of physical sciences of the time. 
Blokhintsev was certainly influenced strongly by
Mandelstam and learned a lot from him, in
particular, his breadth of views on physics as an indivisible
science, lecturing skills, understanding the importance
of a scientific school, organization of science, etc.
As was noted later, ''Lectures and seminars given by
Mandelstam at the university in 1925-1944 were of
special importance. They were devoted to a wide field
of the most topical problems in physics in which the
lecturer delivered an extremely deep analysis of the
modern state of the art without concealing existing difficulties,
and he outlined original solutions to very complex
problems. These lectures attracted a wide audience
of physicists of various ages and ranks from all parts of
Moscow.'' Mandelstam delivered his famous lectures
on the principles of quantum mechanics (the theory of
indirect measurements) in spring of 1939. He
intended to read a series of lectures on the connection
of the mathematical tools of quantum mechanics and its
statistical interpretation, causality, etc., as a continuation
of these lectures; the basis of this series of lectures
was supposed to be the famous book written by J. von Neumann.
Later, this program was realized by Blokhintsev.\\ 
It was time  when quantum mechanics had acquired
a certain maturity~\cite{jam}. 
In the book by Gurney~\cite{gur}, also
referred to in Blokhintsev's works, quantum mechanics
is characterized as a new language of physics and
chemistry. ''The program of quantum mechanics
includes no more and no less than the reconsideration
of atomic and molecular physics in their entirety on the
basis of new laws of behavior of particles following
from quantum mechanics''. 
Blokhintsev joined the realization of a
program of reconsideration of atomic and solid state physics in their entirety on the basis of new quantum physics with
enthusiasm. As he later recollected, ''During that period (1927-1929), new quantum mechanics originated and great capabilities in the 
application of this new physical concept and new methods of calculation of various atomic phenomena were found''~\cite{dib1}.
At that time, solid state physics, in particular, the theory of metals, attracted great attention. In 1932, the work
''Temperature Dependence of the Photoeffect on Pure Metals'' of D.I. Blokhintsev was published. The next paper was
''The Work Function of Electrons from Metals'' (1933) (jointly with I.E. Tamm). In the monograph~\cite{mott} this study by Tamm and
Blokhintsev was cited together with other basic works on the problem. Thus, from the very beginning, his works were
at the highest level of quality. 
The early works of D.I. Blokhintsev have manifested also his talents of clear vivid presentation of the subject, 
transparent style, concreteness, the ability to point out most significant things and, most important, emphasis on the
\emph{physical meaning}. In a large work by Blokhintsev in 1933 ''Theory of Electron Motion in a Crystal Lattice'', the 
F.Bloch theory of motion of tight binding electrons was generalized for the  many bands case and for the electron
motion in a crystal which is bounded by surface. The next work was the paper ''Theory of Anomalous Magnetic and Thermoelectric 
Effects in Metals''(1933) coauthored with L.W. Nordheim (1899-1985). In this work, the consistent theory of thermoelectric and
galvanomagnetic effects in metals was constructed. Unlike earlier works, the case of $s-p$  band metals was considered.
The authors studied the behavior of divalent metals in a magnetic field ( Thompson and Hall effects). To
make their equations compact, Blokhintsev and Nordheim introduced a new notion, the tensor of reciprocal effective masses.
In the book of Mott and Jones~\cite{mott},  the priority of Blokhintsev and Nordheim in the  creation of 
this fundamental notion was established. The achievement made by Blokhintsev and Nordheim was that they showed that the concept of effective mass
was much more general and workable than had been assumed before and for the first time demonstrated the tensor character of the
effective mass by considering the behavior of the electron in external fields. 
It turned out that the notion of effective mass is extremely useful in the 
theory of conductivity and other fields of solid state physics, nuclear physics, etc.
The concept of effective mass became widely applied,
especially in semiconductor physics and the physics of semiconductor devices, the polaron theory, semiconductor superlattices,
microelectronics and physics of nanostructures.\\
A few word should be said about Blokhintsev's coauthor
Lotar Wolfgang Nordheim (1899-1985). Nordheim
belonged to the Gettingen school of theoretical
physics. He was a PhD student with M. Born, and after
defending his PhD thesis in 1923, his assistant and colleague
till 1933. All his works are marked by bright talent
and deep insight into a problem. In Jammer's book~\cite{jam}, the following
fact is given: ''In autumn of 1926, Hilbert began
systematic studies of the mathematical principles of
quantum mechanics. Lotar Wolfgang Nordheim,
Born's former student, and the 23-year-old John von
Neumann helped him in these studies. Hilbert also gave
lectures on the mathematical principles of quantum theory,
which were published in shorter form in the
spring of 1927.''
Nordheim worked successfully in the application of
quantum mechanics to statistical physics and solid state
physics. He gave a successful description of the electron
work function in metals, thermoelectron emission,
electron kinetics in metals and alloys, etc. 
Thanks to a grant from the Rockfeller Foundation,
Nordheim visited Moscow in 1933 as an invited professor
to MSU. His studies were quite close to those performed
by the Tamm's group. It was during that visit
that he performed his joint work with Blokhintsev.\\
In 1933,  Blokhintsev published ''Theory of the Stark Effects in a 
Time-Dependent Field''. In this paper Blokhintsev showed that the atomic levels move under influence of  variable electric 
field (Stark modulation). The picture of light scattering depends nonlinearly on the intensity of the incident light. This work
was one of the first in the field of physics, which was latter called \emph{nonlinear optics}.\\In 1934,  Blokhintsev 
published paper on the theory of phosphorescence. According to the author, ''An attempt was made to explain the phenomenon
of phosphorescence in the so called Lenard phosphors on the basis of quantum mechanical ideas of the electron motion in the 
crystal lattice''~\cite{dib1}. Blokhintsev assumed that duration of the phosphorescence can be related with the capability of formation of
\emph{quasilocalized} electronic states in a real crystal as a result of the \emph{local lattice deformation} due to the
introduction impurities. Then he estimated the time of reciprocal recombination of these states. Thus, the theory of localized
states made it possible to qualitatively ( and, partially, quantitatively) interpret the big duration of the phosphorescence.
This point of view was included in textbooks on optics. This and subsequent works by Blokhintsev, in which the detailed theory
of the kinetics of phosphorescence in heteropolar crystals and the theory of dyed crystals were constructed, contributed
considerably to deeper understanding of this problem and showed once more that the quantum mechanical approach is indeed the
''new language of physics and chemistry'', providing effective description of phenomena considered ''mysterious'' in classical
physics. The same approach was used by Blokhintsev in the work ''Quantum Mechanical Theory of Adsorption'' (1934) (co-authored with
Sh. Shekhter). This work is a very useful and clear survey of the problem as a whole. The paper of the same authors ''Lifetime of
Particles in Adsorbed State'' (1934) was devoted to the calculation of the lifetime of particles in the adsorbed  state. 
In that
paper it was demonstrated how the quantum mechanics provides one with the microscopic picture of phenomenon. The authors obtained the 
correct qualitative behavior of the average lifetime of the adsorbed molecule on the surface, which demonstrated once more the
effectiveness of the quantum mechanical approach. In  1934, Blokhintsev presented his Ph.D. thesis to the Institute of 
Physics of the Moscow State University, entitled \emph{Selected Problems of the Solid State Theory, Especially Metals}. As a result
of the high level of the work, he received a degree of Doctor of Science. At the time, Blokhintsev was 26 years old.\\
In 1935--1936, Blokhintsev continued his work on the theory of light absorption in heteropolar crystals, the theory of 
phosphorescence, and the theory of dyed crystals. It is interesting to note that in the paper ''Theory of Dyed Crystals''
(1936), Blokhintsev, in certain sense, anticipated the concept of the \emph{polaron}, which was formulated later by
S.I. Pekar (1917-1985). S.I. Pekar  wrote this story in his well known monograph~\cite{pek} in 1951: ''In 1936, Blokhintsev attempted
to find out in which crystals autolocalization of electrons pointed out by Landau should be expected on the basis of the
approximation of tight-binding electrons...''. As is well known, S.I. Pekar coined the very term, \emph{polaron}, in 1946.
The main idea was that ''excess'' electron in ionic crystal polarizes the crystal lattice; this polarization in turn
influences the electron, and this action is equivalent to the action of some effective potential well. The depth of this well in 
some crystals may be sufficiently large for discrete energy levels to exist in it. Local polarization caused by the electron 
is related to the displacement of ions from their average equilibrium positions. These states of the crystal   with the 
polarization well in which the electron is localized were termed \emph{polarons} by Pekar. The contribution made by
Blokhintsev in 1936 to this direction of researches was mentioned later by a few other investigators. The main point was
the formulation of the problem of autolocalized electronic states on the basis of approximation of tight-binding electrons.
This approximation (LCAO)~\cite{kuzrev} later become widely used in condensed matter physics, especially for the
description of localized states of different nature and disordered systems. The investigation of localized states in the
framework of the tight-binding approximation bringed Blokhintsev to the point, namely to the need to describe  the interaction
of the electron with the lattice vibrations accordingly to the spirit of tight-binding approximation. This was carried out
 much later (see for details Ref.~\cite{kuzrev}).\\
In 1938, Blokhintsev  prepared his work ''\emph{The
Shift of Spectral Lines Caused by the Inverse Action of
a Radiation Field}'' for publication. He presented it at a
seminar of the Physical Institute of the Academy of Sciences
of the USSR, where he was employed; he also
submitted it to Zhurnal Experimental'noi i Teoreticheskoi
Fiziki [Journal of Experimental and Theoretical
Physics] (ZhTEF). The work was rejected by the
editorial board and published only in 1958 in Dubna in
a collection of Blokhintsev's scientific works and
papers. This work was mentioned in the survey report
delivered by Ya.A. Smorodinskii~\cite{smor} in 1949. Later on, the following
was written~\cite{sborn}: ''Already in early works by Blokhintsev, deep understanding of the essence of quantum
mechanics, fresh and bold ideas, an original way of
thinking that foreshadowed the further development of
physics were evident. Typical in this respect was his
work on the calculation of the 'shift of spectral lines
caused by inverse action of a radiation field,' which in
essence contained the theory of the Lamb shift, which
was the beginning of quantum electrodynamics. It was
reported at the seminar at the Physics Institute of the
Academy of Sciences of the USSR and submitted to
ZhTEF in 1938. The formula for the Lamb shift
obtained by Blokhintsev became famous; it differs
from the Bethe formula only by the numerical factor
added in 1948 as a result of ultraviolet cutoff. Unfortunately,
this important discovery was not published at
that time in ZhTEF. There were no other outlets for
publication''.
The genesis of the work ''The Shift of Spectral Lines
Caused by the Inverse Action of a Radiation Field'' was
best described by Blokhintsev himself~\cite{dib1}. ''I delivered
the work that, in essence, contained the theory of the
Lamb shift discovered ten years later, at a seminar at the
Physics Institute. However, my work was not published,
since the editorial board of ZhETF returned the
manuscript because the calculations were considered
unusual. I kept the manuscript, which was stamped by
the journal certifying the date of its receipt (February 25,
1938). I found no support from my colleagues at the
Physics Institute. There were no other outlets. Thus,
this important work was not published in due time. The
main idea of the work followed from my deep belief
that a physical vacuum existed in reality; however, I
refrained from presenting the affair in this light...''.
The \textbf{Lamb shift} is indeed related to quite remarkable
and interesting effects of quantum physics~\cite{trigg}.
Lamb and his colleagues performed very precise, thorough,
and elegant experimental studies   on the
determination of the structure of levels with $n = 2$ for
hydrogen, deuterium, and singly ionized helium. Since
the energy difference for these levels is very small, the
probability of spontaneous transitions turns out to be
negligible. However, if the atom is placed in a rotating
(or oscillating) magnetic field with the corresponding
frequency, induced transition can be observed. This frequency
can be exactly measured; it is equal to the difference
in energies of the two levels divided by $\hbar$. The
measurement of the rotation frequency in Lamb's
experiments yielded a value of the energy difference of
two levels with the same principal quantum number in
Rydberg units; it is interesting that this does not require
any preliminary data on the Planck constant $\hbar$. The
Lamb shift is mainly determined by the variation in the
''scale'' in wave functions of the atom, which are used
upon calculation of the mathematical expectation of
corresponding expressions. 
Blokhintsev wrote about his calculations in~\cite{dib1}:
''As a result of them, the following expression is
obtained for the frequency shift:
\begin{equation}\label{ls3}
\delta \omega_{0} = k (\frac{e^{2}}{\hbar c})^{3} \frac{Z^{4}}{n^{3}} R \lg \left( \frac{\mu c^{2}}{\Delta E_{av}}\right), 
\end{equation}
where $k$ is the numerical coefficient,$\Delta E_{av}$ is the average
energy, $n$ is the principal number of the level, and $R$ is
the Rydberg constant. Due to the inaccuracy in cutoff,
the coefficient $k$ and the values of $\Delta E_{av}$ differ somewhat
from exact values obtained using the method of electron
mass renormalization (note that (\ref{ls3}) can be rewritten
in the form $\delta \omega_{0} \cong |\psi_{s} (0) |^{2}$, as is usually done now;
here, $\psi_{s} (0)$ is the value of the wave function at the point
$r = 0$). The ratio $\delta \omega_{0}/\omega = 2.8 \cdot 10^{-8}$    calculated using this
formula for the $He$ ion is in good agreement with
respect to its absolute value and sign with the value
measured by Paschen ($10^{-6} - 10^{-7}$). At the time, there
were no more precise measurements. This circumstance
was of course unfavorable for further improvement
of an unpublished work. Only after World War II,
in 1948, did the importance of this work for theoretical
physics become clear.''
The Lamb shift  in levels in hydrogen, i.e., the
energy by which the $2S_{1/2}$ state is higher than the $2P_{1/2}$
state, is obtained by combining different terms contributing
to the theoretical expression for the Lamb shift. Experimental
investigations of the Lamb shift continue.
It was reported not long ago that two-loop corrections
to the Lamb shift were first measured in
strongly ionized atoms of heavy elements using the ion
trap technique~\cite{trap}.
The history of theoretical calculation of the Lamb
shift value is quite interesting. It is known from firsthand
accounts and has been well described in many papers
 and books~\cite{VW,MM}. According to V. Weisskopf ~\cite{VW}, ''Since 1936, there have been vague data
that the position of observed hydrogen levels does not
exactly match the predictions following from the Dirac
equation, the so-called Pasternak effect. Certain considerations
existed on possible ways of calculating this
effect using quantum electrodynamics in the presence
of deviations. After the war, I decided to investigate this
problem together with a very capable \emph{PhD} student,
B. French. We wanted to calculate this effect, which
was more well known as the Lamb shift, by isolating
the infinite self-energy of the electron. These were
complicated calculations, since the renormalization
technique had not been developed yet. It was necessary
to calculate the energy difference of the free and bound
electrons when both energies were infinite. We had to
be very accurate, since the calculation of the difference
of diverging quantities often results in errors. We overcame
difficulties slowly, since there were no good
experimental results at that time. Then Lamb and Retherford
set up a good experiment, and finally, we
obtained a result that agreed well with experimental
data. I informed Julian Schwinger and Dick Feynman;
they repeated the calculations; however, their results
were different from ours, and Schwinger and Feynman
obtained the same number. We postponed publication
to find the error, spending half a year on it. Meanwhile,
Lamb and Kroll published calculation result of the
same effect, which more or less agreed with our result.
Then Feynman called me from Ithaca, ''You were right;
I was wrong!'' Thus, if we had had courage to publish
our results, our paper would have been the first one to
explain the experiment performed by Lamb and Retherford.
What's the moral of this story? You have to
believe in what you do.''\\
In 1939, Blokhintsev published his work ''Hydrodynamics
of an Elecron gas''. In this work, the
hydrodynamic description of the system of many particles
(electrons), i.e., description in terms of a ''reduced''
set of variables characterizing the system, the current $I(x)$
and the particle density $\rho(x)$, was considered. 
Blokhintsev maintained that since a many-particle problem
could not be solved exactly, an approximate solution
should be sought. It is known that an efficient
way for calculating the energy eigenfunctions and
eigenvalues is the self-consistent field method. This
method was first developed by Hartree without taking into
account electron exchange and then by Fock  with
this exchange taken into account. There exist a large
number of works on this method  both with and
without the exchange account. Blokhintsev wrote in
his work that from the very beginning he used the Hartree-~Fock approximation, 
which assigns an individual function
$\psi_{k}(x)$ to each electron $n$. In this approximation, the
system of electrons is described by the density matrix.
 Considering the dynamic
equations (equations of motion) for the current,
Blokhintsev derived the ''\emph{hydrodynamic}'' equation for a
system of many particles (electrons) that contained gas
density gradients in the stress tensor. To obtain closed
expressions, he used approximations characteristic of
statistical Fermi-Thomas theory. It is known  
that the statistical model of the atom describes the electrons
of the atom statistically as an electron gas at a
temperature of absolute zero. The model yields good
approximation only for atoms with a large number of
electrons, although it had been used for up to ten electrons.
For the statistical approach, the details of the
electronic structure had not been described; therefore,
the application of a hydrodynamic description was
quite relevant. Following the spirit of the statistical
model of the atom, the total energy of the atom is
obtained from the energy of the electron gas in separate
elementary volumes $dv$ by integrating over the whole
volume of the atom. Working in this way and using the
continuity equation, Blokhintsev derived an expression
for the gas energy that (in the statistical case) coincided
with the expression obtained earlier by Weizsacker
 using a different method.\\ 
It is appropriate to note here that the work ''Hydrodynamics
of an Electron Gas'' contains one more
aspect that does not seem striking at first sight but is
nonetheless of great interest. In essence, it was shown
in this work that a system in the low-energy limit can be
characterized by a small set of ''collective'' (or hydrodynamic)
variables and equations of motion corresponding
to these variables. Going beyond the framework
of the low-energy region would require the consideration
of plasmon excitations, effects of electron
shell reconstructing, etc. The existence of two
scales, low-energy and high-energy, in the description
of physical phenomena is used in physics, explicitly or
implicitly. Recently, this topic obtained interesting and
deep development, connected with the
concept of the ''\emph{quantum protectorate}.'' In a work with
a remarkable title, ''\emph{The Theory of Everything}''~\cite{pnas},
authors R. Laughlin and D. Pines discussed the most
fundamental principles of the description of matter in a
wide sense. The authors put
forward the question what the ''Theory of Everything''
should be. In their opinion, 
''it describes the everyday world of
human beings - air, water, rocks, fire, people, and so forth''.
The answer given by the authors was that ''this theory is
nonrelativistic quantum mechanics,'' or, more precisely,
the equation of nonrelativistic quantum mechanics,
which they wrote in the form
\begin{equation}
\label{seq}
H\psi = - \frac{\hbar}{ i} \frac {\partial \psi}{\partial t}.
\end{equation}
That was the only formula in their work; they also
gave detailed definition of the Hamiltonian of a system
consisting of many interacting particles. The authors
agreed, however, that 
''Less immediate things in the universe, such as the planet
Jupiter, nuclear fission, the sun, or isotopic abundances of
elements in space are not described by this equation, because
important elements such as gravity and nuclear interactions are
missing. But except for light, which is easily included, and
possibly gravity, these missing parts are irrelevant to people-scale
phenomena. Eq.(\ref{seq}) is, for all practical purposes, the
Theory of Everything for our everyday world.
However, it is obvious glancing through this list that the
Theory of Everything is not even remotely a theory of every
thing. We know this equation (\ref{seq}) is correct because it has been
solved accurately for small numbers of particles (isolated atoms
and small molecules) and found to agree in minute detail with
experiment. However, it cannot be solved accurately when
the number of particles exceeds about 10. No computer existing,
or that will ever exist, can break this barrier because it is a
catastrophe of dimension. If the amount of computer memory
required to represent the quantum wave function of one particle
is $N$ then the amount required to represent the wave function of
$k$ particles is $N^{k}$.'' 
According to R. Laughlin and D. Pines,
''The emergent physical phenomena regulated by higher organizing
principles have a property, namely their insensitivity to
microscopics, that is directly relevant to the broad question of
what is knowable in the deepest sense of the term. The low energy
excitation spectrum of a conventional superconductor,
for example, is completely generic and is characterized by a
handful of parameters that may be determined experimentally
but cannot, in general, be computed from first principles. An
even more trivial example is the low-energy excitation spectrum
of a conventional crystalline insulator, which consists of transverse
and longitudinal sound and nothing else, regardless of
details. It is rather obvious that one does not need to prove the
existence of sound in a solid, for it follows from the existence of
elastic moduli at long length scales, which in turn follows from
the spontaneous breaking of translational and rotational symmetry
characteristic of the crystalline state. Conversely, one
therefore learns little about the atomic structure of a crystalline
solid by measuring its acoustics.
The crystalline state is the simplest known example of a
quantum protectorate, a stable state of matter whose generic
low-energy properties are determined by a higher organizing
principle and nothing else. There are many of these, the classic
prototype being the Landau fermi liquid, the state of matter
represented by conventional metals and normal $^{3}He$... Other important quantum protectorates
include superfluidity in Bose liquids such as $^{4}He$ and
the newly discovered atomic condensates, superconductivity, band insulation, ferromagnetism, 
antiferromagnetism, and the quantum Hall states. The
low-energy excited quantum states of these systems are particles
in exactly the same sense that the electron in the vacuum of
quantum electrodynamics is a particle: They carry momentum,
energy, spin, and charge, scatter off one another according to
simple rules, obey Fermi or Bose statistics depending on their
nature, and in some cases are even ''relativistic,'' in the sense of
being described quantitatively by Dirac or Klein-Gordon equations
at low energy scales. Yet they are not elementary, and, as in the
case of sound, simply do not exist outside the context of the
stable state of matter in which they live. These quantum protectorates,
with their associated emergent behavior, provide us
with explicit demonstrations that the underlying microscopic
theory can easily have no measurable consequences whatsoever
at low energies. The nature of the underlying theory is unknowable
until one raises the energy scale sufficiently to escape
protection.''\\ The existence of two scales, low-energy and the
high-energy, in the description of magnetic phenomena
was stressed by Kuzemsky (see Refs.~\cite{kuz81,akuz00,kuze02}) upon comparative investigation
of localized and itinerant quantum models of magnetism. The concept of quantum
protectorate was applied to the theory of magnetism
in paper~\cite{kuze02}. We succeeded in formulating the criterion
of applicability of quantum models of magnetism
to particular substances on the basis of analyzing their
low-energy and high-energy spectra.\\
In 1940, Blokhintsev's attention was attracted to the
problem of statistical description of quantum systems.
Interest to these problems stemmed from lectures and
works on quantum mechanics by L. I. Mandelstam and
K.V. Nikol'skii. Nikol'skii's book \emph{Quantum Processes}~\cite{nik}
is mentioned many times in his papers.
In the work ''Correlation of a Quantum Ensemble
with a Classical Gibbs Ensemble'' (1940), the limiting
transition from quantum equations of motion for the
density matrix to the equations of motion for the classical
distribution function was studied. 
Blokhintsev studied the possibility of correspondence
between the classical distribution function $f(q, p)$ and
the quantum density matrix $\rho$ from the general point of
view. For this purpose, the mixed $(q, p)$ representation
for the density matrix  was used. 
  Blokhintsev shown in that paper  
that there does not exist any distribution function
depending on $(q,p)$ which could describe the quantum
ensemble.
In the next work on the topic (1940), the problem of
the conditions of approximation of quantum statistics
by classical statistics was considered. It was shown that
there is no limiting transition $(h \rightarrow 0)$ from a quantum
ensemble consisting of similar particles to a classical
ensemble. The classical description is obtained if the
state of the system is characterized by the position in
the phase cell $\Omega \gg \hbar$. Thus, in these works, a new direction
of physics was initiated: quantum mechanics in the
phase space~\cite{qmphs}.\\The title of the next work written by Blokhintsev
 (jointly with Ya.B. Dashevskii in 1941) is ''Partition
of a System into Quantum and Classical Parts.''
According to the authors, ''Among physical problems
that should be solved using quantum mechanical methods,
there are such problems in which the system of
interacting particles under study has a property that one
of its parts during the processes occurring in the system
moves as though it obeys classical laws of motion, i.e.,
moving along a trajectory.''
In this work, they studied the possibility of partitioning
an interacting system into quantum and classical
parts. They demonstrated the type of perturbation when
the classical part acts on the quantum part. 
This field attracted great interest in subsequent years,
especially in many problems of physical chemistry. A
large number of works are devoted to this topic; some of
them are considered in detail in survey~\cite{kap}.\\
In 1946, after switching to defense problems,
Blokhintsev returned to quantum physics. The work
performed in 1946 is titled ''Calculation of the Natural
Width of Spectral Lines Using a Stationary Method''.
 This short work demonstrated high flexibility in
handling tools of quantum mechanics when the result
was reached in a simple and elegant way. Blokhintsev
wrote, ''Usually the problem of emission and absorption
of light is considered using the method of quantum
transitions. However, this problem, similar to the dispersion
problem, can be solved in an extremely simple
way using the method of stationary states''. Then,
the author wrote out the system of equations for state
amplitudes of two types: (a) when the emitter is in the
state $m$ and light photons are absent, and (b) when the
emitter is in the state $n$ and one light photon has been
emitted. Taking into account the energy conservation
law, the solution for the amplitude was obtained, and on
its basis, the approximate expression for the level position
of the whole system (emitter and radiation). ''This
expression resulted in exactly the same shift and smearing
of levels as those obtained by Dirac upon calculation
of resonance scattering.'' Then, the spectral distribution within the line width
was found. The author noted that upon transformation
of the amplitude to the coordinate representation, , ''we
obtain a divergent wave with an amplitude that slowly
increases with increasing distance from the radiation
source in the same way as took place for a classical
decaying oscillator''.\\
In 1947, Blokhintsev published the work ''The Atom
under an Electron Microscope''.
Blokhintsev wrote that ''this work, devoted to a very
special problem, is worth mentioning due to a somewhat
unusual formulation of the problem. The origin is
thus. I paid attention to the fact that under the action of
a scattered electron, the atom receives recoil and can be
knocked out of its position on the surface of the 'object
plate.' If it were not knocked out at first scattering, it
could be knocked out at subsequent scattering. It
should be noted that this experiment is unusual from the
point of view of the common formulation of measurements
in a quantum ensemble. Indeed, in this case, we
consider the repetition of measurements with the same
sample of the atom, rather than a set of atoms, as is usually
done. After each measurement the state of the
atom, generally speaking, changes, and it becomes a
sample of another quantum ensemble. Thus, the series
of scattering necessary for obtaining an image of the
atom consists of a series of scattering related to objects
from different quantum ensembles. This seems to be a
unique case of such a situation.''\\
Since physicists, chemists, metallurgists, and biologists
needed improved microscopes, this problem
always stirred interest. It should be noted that remarkable works were performed
by Mandelstam on the theory of the microscope.
 Mandelstam displayed his inherent the
strength and depth of thought and his keen understanding
of the physical nature in analyzing this problem.
Blokhintsev's work continued the development
of the theory of the microscope at the new quantum
stage. The interest in this problem not only stemmed
from the applied value. According to Blokhintsev, ''The
development of the theory of the microscope is of interest
from the theoretical point of view, since when
observing a single atom using an electron microscope,
the image will emerge as a result of repetition of single
scattering acts on the same object, while in quantum
mechanics, results are usually formulated with respect
to a set of objects in the same initial state. Due to the
action on the atom, each new scattering act, generally
speaking, will force the atom to be in a new initial state.
Therefore, it is important to analyze the influence of
electron scattering on the state of the observed atom''.
Further development in physics proved that Mandelstam
and Blokhintsev's interest in problems of the theory
of the microscope was justified. This direction was
developed in subsequent years greatly and is
being extensively developed now.\\
Blokhintsev's name is closely related to the problem
of \textbf{interpretation of the quantum mechanics}~\cite{jam2}.
Blokhintsev recollected~\cite{dib1} that ''in the 1930s --
1940s, the interest of many physicists-theoreticians at
the Lebedev Physics Institute and MSU was concentrated
on the principles of quantum mechanics, which
seemed full of paradoxes to many people.''
 A large number of
his books~\cite{dib2,dib3,dib4} and papers~\cite{dib5,dib6}   were
devoted to this problem. His views of the problem
changed and evolved with deepening and perfection of
arguments.\\ 
The most
topical problems of interpreting quantum mechanics
were the problem of measurement and the role of the
observer, and the probabilistic interpretation of the wave
function.  
The variety of opinions concerning
the interpretation of quantum mechanics
increased with time. Blokhintsev wrote~\cite{dib1}: ''Those
discussions are reflected in my works; the polemical
character of my papers devoted to critical analysis of
the ideas of the Copenhagen school and those of Fock
gradually brought me to a consistent materialistic concept
of quantum ensembles and mathematical measurement
theory. Only in the 1960s, after discussions with
the Hungarian physicist L. Janosi, did I manage to formulate
a reasonable theory of quantum measurements
free from inconsistencies in interpreting the role of the
observer. In this new concept, the measuring device and
its interaction with the microobject were transformed
from the subject of philosophical discussions to the
subject of theoretical physics''.

 As a result of longterm
research and reflections, Blokhintsev developed
his own approach to interpreting quantum mechanics,
which included ideas put forward by J. von Neumann,
L.I. Mandelstam, and K.V. Nikol'skii. It was called
the interpretation of quantum mechanics on
the basis of quantum ensembles.
He wrote in a
summary work~\cite{dib4}, ''The presentation of quantum
mechanics undertaken in these lectures is essentially
based on the ideas of von Neumann, which in their time
attracted the attention of the Moscow school of theoreticians;
in 1930s this school was headed by Academician
Mandelstam; also Nikol'skii contributed considerably
to our understanding of quantum mechanics.'' Blokhintsev
thought that ''this approach to the principles of
quantum mechanics had an advantage, as compared to
traditional interpretations on the basis of the wave function,
since it allowed one to include the theory of quantum
measurements as a chapter of quantum mechanics''.
In Blokhintsev's approach, the statistical operator
describing the state of the microsystem in a quantum
ensemble of the general type plays the primary role.
The wave function describes a special type of quantum
ensemble, the coherent ensemble. 
Blokhintsev's approach to the interpretation of
quantum mechanics became widely known.  De Witt and Graham~\cite{bdw2}
 in their survey of different approaches to
interpreting quantum mechanics wrote about Blokhintsev's
books: ''...they are both  very well written
and informative. The departure from orthodoxy
occurs, in fact,  only in certain attitudes and choice of words, while
the general presentation of quantum mechanics is refreshing...;
[the second book] contains an excellent account of measurement theory''.
Blokhintsev's approach to interpreting quantum
mechanics is a constituent part of the scope of ideas of
various researchers. One of the authoritative historians
of quantum mechanics, Hooker~\cite{chok}, noted that
''...Einstein and his co-workers  Podolsky and Rosen, Blokhintsev,
Bopp, de Broglie, Popper, Schrodinger, Lande, and
most recently Ballentine   constitute a small
group of physicists and philosophers, who are determined to treat
quantum theory as a species of statistical
mechanics, many of them   hoping ultimately to reinstate
 the classical conception of reality.'' A detailed survey
of the interpretation of quantum mechanics on the
basis of quantum ensembles can be found in~\cite{home}.
The interpretation of quantum mechanics on the
basis of quantum ensembles is one of many in existence.
Thus, the interpretation of quantum mechanics on
the basis of quantum ensembles occupies a separate
(noticeable) place among other possible approaches to
interpretation of quantum mechanics.
Interpretation of quantum mechanics on the basis of
quantum ensembles was considered in detail by
Ya.A. Smorodinskii (1986). The conclusion he made is
quite remarkable~\cite{smor2}: ''Discussion showed that if the theory
of quantum ensembles is used, these ensembles should
be assigned unusual properties that could not be consistent
with common probability theory; these properties
are not manifested for one particle and can be found
only in correlated effects; similar to non-Euclidean
geometry necessary for the description of the velocity
space in special relativity, quantum mechanics has generated
the \emph{non-Kolmogorov probability} theory; this is
probably the deep meaning of analysis of the properties
of a quantum ensemble'' (see also recent book~\cite{xren}).\\
We conclude this paper with the words by Max Born formulated
in his lecture ''Experiment and Theory in Physics''
delivered in 1943. ''Those who want to master the
art of scientific prediction should, instead of relying on
abstract deduction, try to comprehend the secret language
of Nature, which is represented by experimental
data.'' Blokhintsev in his lectures and talks more than
once expressed similar thoughts, may be in slightly different words.
In these notes and in  the extended review~\cite{kuzrev} we have tried not only to write about
Blokhintsev's studies, but build them into appropriate
lines of the development of quantum physics and connect
them, directly or indirectly, to the modern development
of these fields of science. We have tried to show that
Blokhintsev's book  \emph{Quantum Mechanics}~\cite{dib2},
which is justly considered one of the best textbooks in
quantum physics, was compiled by a witness to and a
participant in the formation and development of quantum
mechanics. It organically includes most of his original
works in an integrated description of the subject.
This, together with the definite literary talent of the
author and his gift for presenting the subject clearly and
lucidly, is the background on which the book  \emph{Quantum Mechanics}
 stands, and it continues to describes the world using
the \textbf{language of a quantum}!\\
In this work due to the
lack of space, not all the topics and problems that
I wanted to discuss are here.  Permit me to
refer any reader who wants to reflect on Blokhintsev's
works to a collection of selected works in two volumes
that will be published in 2008 in Moscow.\\
A more detailed discussion of modern approaches to
interpreting quantum mechanics can be found in paper~\cite{kuzrev}.
The full details and precise References are given in paper~\cite{kuzrev}
as well. 
\end{document}